\documentclass[10 pt]{article}
\usepackage{euscript}
\usepackage{latexsym}
\newcommand{\ESE}{\ensuremath{\EuScript{E} }}

\newcommand{\qv}{\mbox{\boldmath $q$}}  
\newcommand{\rv}{\mbox{\boldmath $r$}}  
\newcommand{\kv}{\mbox{\boldmath $k$}}  
\newcommand{\rholg}{\mbox{\Large $\rho$}}

\newcommand{\lang}{\mbox{$<$}}
\newcommand{\rang}{\mbox{$>$}}
\newcommand{\Uc}{\mbox{ $\EuScript{U}_{c}$}}
\newcommand{\lcommb}{\mbox{\Large \boldmath ${[ }$}}
\newcommand{\rcommb}{\mbox{  \Large \boldmath ${ ]}$}}
\usepackage{graphics}

\begin{document}

\large  

\title{ A Quantum Description of Brownian Motion and the RC Circuit}

\author{W. H. Richardson \\
Qusemde, 101 Industrial Road, Belmont CA 94002 \\
and \\
Department of Applied Physics,  Stanford University, Stanford, CA 94305-4085 \\}

\date{ }

\maketitle

\begin{abstract}
\normalsize

An equation for the reduced density matrix  which describes  a free particle, that is  interacting with a linearly dissipative  medium, is derived  using the total Hamiltonian, and without resorting to  any artificial model.   A Master equation is also obtained for the dissipative medium.   The theory  erases the notion of the reservoir.  It is shown that the  dynamical interaction with the medium  is not completely determined by the   friction constant (or the corresponding fluctuation-dissipation relation).   Unexpectedly, through the density-density correlation, the longitudinal dielectric function is shown to play a critical role. 

\end{abstract}

\newpage

The process of dissipation is   almost the antithesis of unitary evolution of the wavefunction, and  therefore   a quantum description of even the simplest quantum system (that of a free particle)  interacting with a dissipative medium has turned out to be a formidable problem.  In fact, and as is clear from recent publications \cite{t+sipe} -- \cite{kohen+tannor}, there is still no  agreement on the equation of motion for  the reduced density matrix that describes the particle.   The main problem has been the derivation of a suitable Hamiltonian that describes the whole system.    So far, no one has found operators that can simultaneously  a) describe the essential features of the dissipative medium, b) satisfy a  commutation relationship, and c) be used to construct that part of the  Hamiltonian   that describes the interaction with the free particle.    Consequently,  the dissipative medium has often been modeled as a collection of harmonic oscillators (or similar entity ) with  features prescribed by the \mbox{fluctuation-dissipation theorem   \cite{louisell} $---$ \cite{WHR:ijmb}. \hspace*{\fill}$----<$np1$>$}

 Among the many  conditions  on the equation for the reduced density matrix, two of the most difficult to fulfill have been  a) an equation that is  valid when quantum fluctuation is larger than thermal fluctuation, and b) an equation  that ensures  positivity of the density matrix.   It was  demonstrated    that in order to describe quantum Brownian motion the  probability diffusion coefficient  typically features  a time dependence, and  does  not vanish at zero temperature \cite{unruh+zurek}, \cite{hu+paz}.  However those works were criticized because positivity of the density matrix was not guaranteed.  On the other hand, in   works where a term was added to the standard Master equation in order to ensure a positive semidefinite  density matrix \cite{diosi}, \cite{sgao}, \cite{bvacc},   none of  the equations   can really describe quantum Brownian motion because the  diffusion coefficient vanishes at zero temperature, and symptomatically  it is not time-dependent.      As will become clear,  a key fragment of the riddle of Brownian motion, as well as    the shortcomings of all  previous  approaches,  originate in the certainty of  the fluctuation-dissipation relation for the friction coefficient.      In the following discourse, I present for the first time a description of  Brownian motion that does not rely on a collection of harmonic oscillators (or similar) model  for  the lossy medium.  A complete  Hamiltonian is derived and used. 

\noindent   \begin{minipage}[]{2.75 in}
 \hspace*{\parindent}  The particular system of interest --- a current driven RC circuit ---, which is identical to a free particle driven by an external force,  is shown in figure 1.  In the absence of a  resistor, the system  is well described by the charge operator $Q$, and operator $\Phi$  \cite{WHR:ijmb}, which   satisfy  a  commutation relationship $[\Phi,Q] =i\hbar$.  And the contribution  to the total Hamiltonian is given by   $H_{c} = Q^{2}/(2C)$,  
 \end{minipage}
\begin{minipage}[]{3.0 in}
\hspace{0.5 in} \scalebox{0.25}{\includegraphics{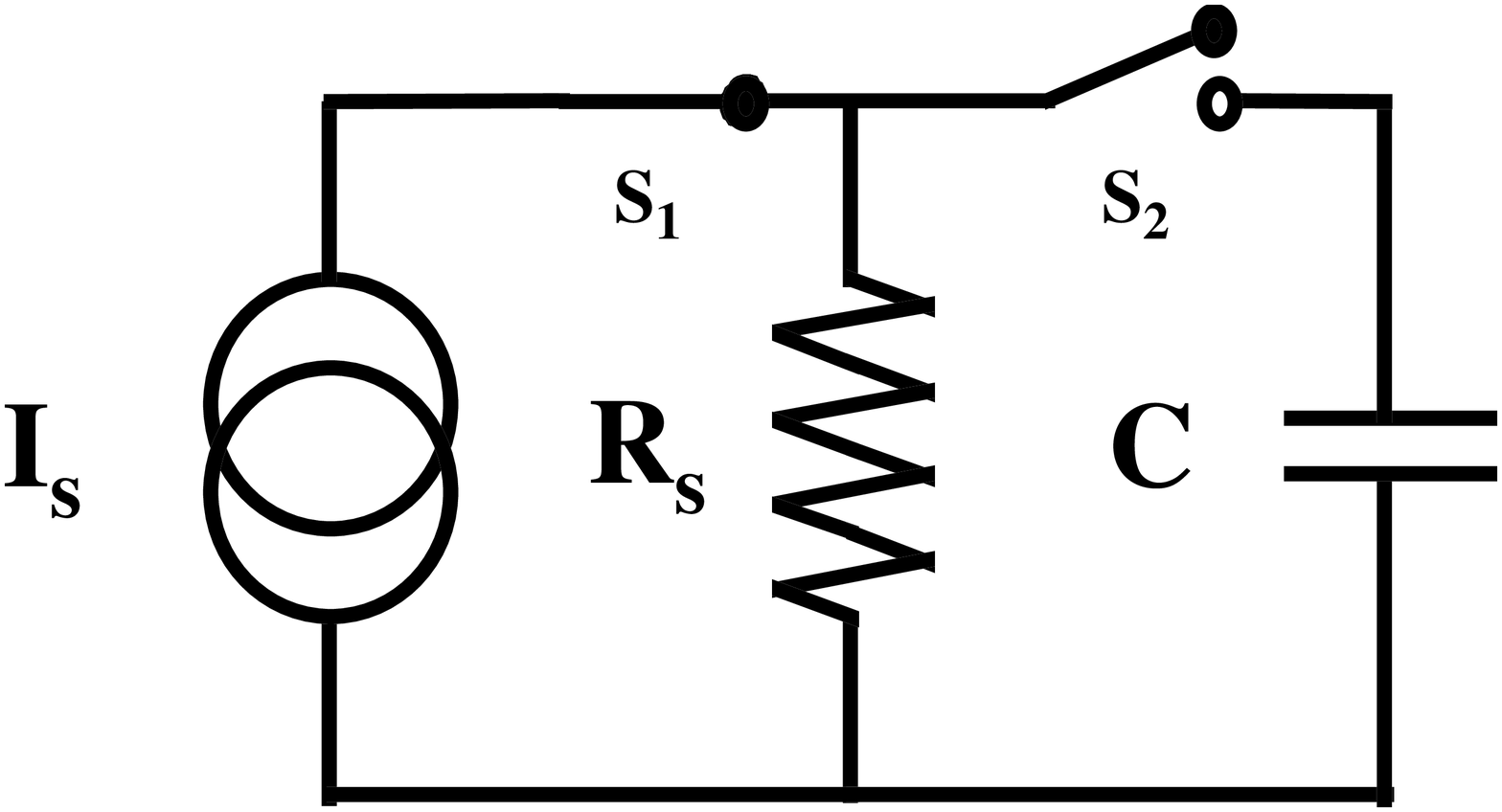}}
 \hspace*{0.5 in} Figure 1: Schematic of the\\ 
 \hspace*{0.5 in} circuit under consideration
\end{minipage}
\\

\noindent plus the energy due to the driving force  $ W_{s} = -I_{s} \  \Phi$.

In this paper,  it is asserted that a   quantum mechanically complete description of the interaction with the dissipative medium  can be  obtained in terms of the intrinsic electrical current operator, \ \ 
$ {\bf J}_{d}(\qv)  =   \sum_{\lambda \eta}  {\bf p}^{\lambda \eta}(\qv) C_{\lambda}^{\dagger} C_{\eta}$ \ , with  \ ${\bf p}^{\lambda \eta}(\qv) =  \sum_{l} \frac{e_{l}}{2mi} \int \, d^{3}r \, e^{ - i \ \qv . \rv }  \{ \phi_{\lambda}^{*}(\rv) \nabla \phi_{ \eta}(\rv) - \phi_{\eta}(\rv) \nabla \phi_{\lambda}^{*}(\rv) \} \ $ ; \ 
and  the charge density operator ($\rholg_{cd}$), \ 
$ \rholg_{cd}(\rv) = \sum_{l} e_{l} \psi_{d}^{\dagger}(\rv) \psi_{d}(\rv) = \sum_{i} \  e_{i} \delta( \rv \  - \  \rv_{i}) $.  Here  $\psi_{d}$ is the wave function of the dissipative medium, \  $\phi_{\eta}$ the basis states, and  $e_{l}$  the charge of the $l^{th}$ specie.   It can  be shown \cite{whr:qbmlg} that in wavevector space  
the operators satisfy a commutation relationship
$ \left [ \ \rholg_{cd} ( \qv ) \, , {\bf J}_{d}( - \qv ) \right ] =  \hbar \qv \sum_{l} e_{l}^{2}/m $ \ ,  
were \ \ $  \rholg_{cd}(\qv) =  \int d^{3} r \rholg_{cd}(\rv)e^{ - i \ \qv . \rv }$ \ . 
  \hspace*{\fill}$-----<$np2$>$

Construction of the Hamiltonian for the complete system started from  the Hamiltonian for the  current driven capacitor, and the interaction energies were derived using   standard electromagnetic theory.  The Hamiltonian was found to be
  \begin{equation}  H = H_{c} + H_{d} + W_{s} + H_{je} + H_{cd} + W_{d} \ \ .
\end{equation}
Here  $H_{cd}$ describes  the interaction between the scalar potential $Q/C$ and the charge density operator : \ $ H_{cd} = - \int d^{3} r \rholg_{cd}(r) Q(r)/C .$ \  
And the energy $H_{je}$ is due to the interaction between the current and field
 $ H_{je} = \int d^{3} r \   {\bf J}_{d}(\rv)  . \  \nabla \Phi(\rv) \ \ $ .
 The interaction energy of the current source driving the resistor is given by the charge density interacting with the equivalent potential $ W_{d} \ = - I_{s}R_{s} \rholg_{cd}(\qv)$.  The   Hamiltonian for  the dissipative medium   is written as:
\  $  H_{d} \ = \ 
\sum_{\kv \sigma} \ESE_{\kv} \, C^{\dagger}_{\kv \sigma}  C_{\kv \sigma}
\ + \  V_{ol}^{-1} \sum_{\qv} V_{i}(q) \rholg_{i}(\qv) \rholg_{cd}(\qv)
\ + \ H_{ee} $ \\ 
, where   ($\rholg_{i}(\qv) = \sum_{j} e^{i \qv.R_{j}}$)  is the impurity density,  $V_{i}(q)$ is the potential associated with the scattering, and $H_{ee}$ is the electron-electron interaction.
\hspace{\fill} $---<$np3$>$

 The total density matrix  is written as
$ \rholg_{\Sigma}(t) = S(t)\otimes \rholg_{d}(t) + \rholg_{c}(t) $, and it 
evolves according to the  equation: \ $
i\hbar \frac{ \partial \rholg_{\Sigma} }{ \partial t} = [ H , \rholg_{\Sigma} ] $.  Here $S(t)$ is the reduced  density matrix that describes the free particle 
 \ $ S(t) = Tr_{d} \ \rholg_{\Sigma}(t) $, \ 
$\rholg_{d}$ describes the dissipative medium \ 
$  \rholg_{d} = Tr_{s} \  \rholg_{\Sigma}(t) $, \ 
and $\rholg_{c}$ is the correlation between the sub-systems. In general, any density matrix can  be written in the above form.
The calculation is done in the interaction representation.  Accordingly  I write
$ H  = H_{0} + W +H_{i} $ \  ,  with $H_{0}  = H_{c} + H_{d}$, and $W=W_{s}+W_{d}$,  and of course  for any operator $O$, \ 
$ \tilde{O}(t) = e^{\frac{i}{\hbar} H_{0}t} O  e^{-\frac{i}{\hbar} H_{0}t} 
$.  It then follows from Liouville's equation that
\begin{equation}
i\hbar \frac{ \partial \tilde{S} }{ \partial t} =  \lcommb \tilde{W_{s}} + \tilde{V_{s}}(t) , \tilde{S}(t) 
\rcommb 
 + Tr_{d}  \lcommb \tilde{W}(t) +  \tilde{H_{i}}(t), \tilde{\rholg_{c}}(t) \rcommb  
\label{S2}
\end{equation}
\begin{equation}
i\hbar \frac{ \partial \tilde{\rholg_{d} } }{ \partial t} =    \lcommb \tilde{W_{d}} + \tilde{V_{d}}(t) ,  \tilde{\rholg_{d}}(t)  \rcommb  + 
Tr_{s}  \lcommb \tilde{W}(t) +  \tilde{H_{i}}(t), \tilde{\rholg_{c}}(t) \rcommb
\label{rhod2}
\end{equation}
\begin{eqnarray}
\label{cdm1}
i\hbar \frac{ \partial \tilde{\rholg_{c} } }{ \partial t}  & = &  -  [ \tilde{V_{s}}(t) + \tilde{W_{s}},  \tilde{S}(t) ]\otimes \tilde{\rholg_{d}}  - \tilde{S} \otimes  [ \tilde{V_{d}}(t) + \tilde{W_{d}} ,  \tilde{\rholg_{d}} ]   + 
[\tilde{W} +  \tilde{H_{i}}, \tilde{S} \otimes \tilde{\rholg_{d}} ]  \\
 & &+ [\tilde{W} +  \tilde{H_{i}},  \tilde{\rholg_{c}} ]
- Tr_{d} \{  [ \tilde{W} +  \tilde{H_{i}}, \tilde{\rholg_{c}}  ] \} \otimes \tilde{\rholg_{d}} - \tilde{S} \otimes Tr_{s} 
[\tilde{W} +  \tilde{H_{i}}, \tilde{\rholg_{c}} ] \nonumber
\end{eqnarray}
and \ $ \tilde{V_{s} } (t) =  Tr_{d} \{ (\tilde{W_{d}}(t) +  \tilde{H_{i}}(t)) \tilde{\rholg_{d}}(t)  \} $ , \ with  $\tilde{V_{d}}(t) =  Tr_{s} \{ (\tilde{W_{s}}(t) +  \tilde{H_{i}}(t)) \tilde{S}(t) \} $ . 

A detailed description of the iterative process by which these  
 coupled  equations (~\ref{S2}),  (~\ref{rhod2})  and (~\ref{cdm1}) were   solved 
will be described elsewhere \cite{whr:qbmlg}.  First  a  solution for  $\rholg_{c}$  was obtained (also via an iterative process),  then that solution was substituted into  (~\ref{S2}); $\rholg_{d}$ was computed using the improved $\rholg_{c}$, and  $S(t)$.  In computing $S(t)$, the initial values were obtained by assuming that initially S1 and S2 are open (see fig.1), and  at t=0 the switches are closed, then the current turned on to a value of $I_{s}$.     It was found that 
\begin{equation}
\label{S4}
 \frac{ \partial S}{ \partial t} =  \frac{1}{i \hbar} [ H_{c} + W_{s} ,  S(t)]
  \ +    \ \Uc \left ( \frac{ \partial \tilde{S_{a}} }{ \partial t}
+  \frac{\partial \tilde{S_{b} }}{ \partial t} \right ) \Uc  
\end{equation}
with  $ \Uc = \mbox{exp}( -i H_{c} t/\hbar)$  and, 
\begin{eqnarray}
\label{S4a}
\Uc \dot{ \tilde{S}}_{a}  \Uc \ \hbar^{2}  & = &  - [ \Phi , \Phi S] \ W_{J}^{+}   +  [ \Phi, Q S] \ W_{J1}^{+} \frac{1}{C}  -   [Q, Q S] \  W_{\rho}^{+} \frac{1}{C^{2}} \nonumber \\  
 & &   + \ [ \Phi , S \Phi] \  W_{J}^{-}   
 -  [ \Phi, S Q ] \ W_{J1}^{-} \frac{1}{C} +  [ Q , S Q ] \  W_{\rho}^{-}\frac{1}{C^{2}} 
\nonumber
\end{eqnarray}
$$
\label{S4b}
\Uc \dot{ \tilde{S}}_{b}  \Uc \ \hbar^{2}   =   -  ([\Phi , [Q , S ]] \ + \ [ Q , [ \Phi , S ]] ) \frac{1}{C} \mbox{Re}( W_{\rho J}^{+})  \  -  \   ([ Q , \{ \Phi , S \}] 
 \  - \  [\Phi , \{ Q , S \}] ) \frac{i}{C} \mbox{Im}(W_{\rho J}^{+}) 
$$

The spectral densities of operators of the dissipative medium, and functions related to the spectral densities appear in equation (~\ref{S4a}). They are defined as follows: \
$ \label{wj+}
W_{J}^{+}(T,t) =   \int_{0}^{t} Tr_{d}  \{ \rholg_{d}(\tau) J^{\dagger}_{d}(\qv,\tau) J_{d}(\qv, 0) \,  \}  L_{0}^{-2} \, d \tau  $ \ ; \ 
$ W_{J1}^{+}(T,t) = \int_{0}^{t} \tau \, Tr_{d} \{ \rholg_{d}(\tau) J^{\dagger}_{d}(\qv,\tau) J_{d}(\qv, 0) \} L_{0}^{-2} \, d \tau $ \ ; \   
$ W_{\rho}^{+}(T,t) = \int_{0}^{t} Tr_{d}  \{ \rholg_{d}(\tau) \rholg_{cd}(\qv,\tau) \rholg_{cd}(-\qv,0) \} \, d \tau   $ \ .  
Other integrals of ensemble averages are similarly defined:  $ W_{J}^{-}(T,t) =   \  \int_{0}^{t} < J^{\dagger}_{d}(\qv,0) J_{d}(\qv, \tau)> d \tau \ L_{0}^{-2}
 $  \ ; \ $ W_{J1}^{-}(T,t) =   \int_{0}^{t} \tau < 
J^{\dagger}_{d}(\qv,0) J_{d}(\qv, \tau)> d \tau \ L_{0}^{-2}$ \ ; \ 
 and \hspace{0.5 cm}  $ W_{\rho}^{-}(T,t) =   \int_{0}^{t} < \rholg_{cd}(-\qv,0) \rholg_{cd}(\qv,\tau)> d \tau .$   The integral of the cross-correlation function 
 is given by  \  $ W_{\rho J}^{+}(T,t) = \int_{0}^{t}  < \rholg_{cd}(\qv,\tau) J_{d}(\qv,0)> d \tau  $.

In the second step,  the solution for $S(t)$  is used in the computation of 
$\rholg_{d}(t)$. Hence now   $\tilde{V}_{d}(t) = Q_{d} Tr_{s}(Q S(t))/C
+ (J_{d} - I_{s}) Tr_{s}(\Phi S(t))$, and from (~\ref{rhod2})  \  it follows that
\begin{eqnarray}
\label{drdt1}
i\hbar \ \frac{ \partial \tilde{\rholg_{d} } }{ \partial t}  & =  &    \left [ \tilde{V_{d}}(t) + \tilde{W_{d}}(t) ,  \tilde{\rholg_{d}}(t) \right ]  +   \frac{1}{i \hbar} \int_{0}^{t} dt' \  Tr_{s} \lcommb  \tilde{H_{i}}(t), [ \tilde{H_{i}}(t') + \tilde{W_{d}}(t'), \tilde{S}(t') \otimes \tilde{\rholg_{d}(t')} ] \rcommb  \nonumber \\  
&  & - \  \frac{1}{i \hbar} \int_{0}^{t} dt' \  Tr_{s} \lcommb  \tilde{H_{i}}(t), \tilde{S}(t') \otimes [ \tilde{V_{d}}(t') ,  \tilde{\rholg_{d}(t')} ] \rcommb  \ \ \ \ . 
\end{eqnarray}

The   spectral density of $J_{d}$  and related  functions were  evaluated  from  the  relationship between the conductivity and the  retarded current-current correlation ( $\Pi_{\alpha \beta}$) function \cite{mahan}:
\begin{equation}
\label{fdth1}
\mbox{Re}\  \sigma_{\alpha \beta}( \qv, \omega)  = - \ \mbox{Im}  \ \Pi_{\alpha \beta}(\qv , \omega)/\omega
\end{equation}
$$
\label{kubo1}
\Pi_{\alpha \beta}(\qv ,\omega) = - \frac{i}{V_{ol}} \int_{ -\infty}^{\infty}
dt \  \theta(t-t') e^{ i \omega (t-t')}  \ < \psi_{d}| \ [ J_{d \alpha}^{\dagger}(\qv,t) , J_{d \beta}(\qv, t') ] \  | \psi_{d} >
$$
where $V_{ol}= L_{0}^{3}$ \ is the volume.
The  trace was  taken with ($ \rholg_{d}(t) = \rholg_{d}(0)$), and the ensemble averages were taken in the long wavelength limit ($\qv \rightarrow 0$).

The spectral density of the charge fluctuation ($W_{\rho}^{\pm}$) in the resistor plays a critical role in determining the off diagonal elements of the density matrix.  This charge fluctuation  is the fluctuation of the conjugate of the intrinsic current operator, and  is not the charge fluctuation on the capacitor. The latter   is determined mainly  by the current fluctuation.   Now the density-density correlation function is related to the {\em longitudinal} dielectric function \cite{mahan}
\begin{equation}
\label{dencor1}
\frac{1}{\epsilon_{\parallel}(\qv , \omega)}  \ = \  1 \ - \ \lim_{ i\omega_{n} \  \rightarrow \ \omega + i\delta} \ \frac{ 4 \pi}{ V_{ol} \, |\qv|^{2} } \int_{0}^{\beta}  d \tau e^{ i \omega_{n} \tau}  < \ T_{\tau}  \ \rholg_{cd}( \qv, \tau) \  \rholg_{cd}( - \qv, 0 ) \ >    \ .
\end{equation}
Evaluation of  $ W_{\rho}^{+} (T, t )$  would have been straightforward if the longitudinal f-sum rule ( $\lim_{\qv \rightarrow 0} Im \, \epsilon^{-1}(\qv , \omega) = -\pi \omega_{p} [ \delta(\omega -\omega_{p}) - \delta(\omega +\omega_{p})]/2$) \ could have been utilized.  But such is not the case since the  low frequency variation of the integrand is important (see also (~\ref{dencor2}) below ).  
In the high temperature or semiclassical  ($\hbar \beta \omega << 1$ ) limit we find 
  $ \lim_{\qv \rightarrow 0} < \rholg_{cd}( \qv, \tau) \  \rholg_{cd}( - \qv, 0 )  >  \ = \   \hbar^{2} \beta e^{-t/\tau_{m}}/(4 R_{s}\tau_{m}) \  $,	   
and hence   \ $ W_{\rho}^{+} (T, t )  \ = \  \hbar^{2} \beta (1-e^{-t/\tau_{m}})/(4R_{s}) $.  
Here $\tau_{m}$ is  the momentum relaxation time of the electron.
\hspace{\fill} $---<$np4$>$

Hence,  the Master equation for $ S(t)$   can be written as
\begin{eqnarray}
\label{ms1}
 \frac{ \partial S}{ \partial t} &  =  &  \frac{1}{i \hbar} [ \frac{1}{2C} Q^{2}  -  I_{s} \Phi ,  S(t)]   \  -  \ 
 [ \Phi ,  [ \Phi , S] ] \ \frac{1}{\hbar^{2}} \mbox{ Re }( W_{J}^{+} ) 
\ + \ [ \Phi , \{ Q , S \} ] \  \frac{i}{\hbar^{2} C } \mbox{ Im} ( W_{J1}^{+} )
  \nonumber \\ 
  &  & \ - \ [ Q ,  [ Q , S] ] \ \frac{1}{(\hbar C)^{2}} \mbox{ Re}(W_{\rho}^{+}) 
\ \ \ + \ \ \  [ \Phi ,  [ Q , S] ] \ \frac{1}{\hbar^{2}C} \mbox{ Re}( W_{J1}^{+} ) 
\ \ \ \ \ \nonumber  \\
&  & \ - \ [ \Phi , \{ \Phi , S \} ] \  \frac{i}{\hbar^{2}} \mbox{ Im} ( W_{J}^{+}) 
\ - \ [ Q , \{Q , S \} ] \ \frac{i}{(\hbar C)^{2}} \mbox{ Im}(W_{\rho}^{+})  \ \nonumber \\
&  & \ - \ ( \hbar S  +  i(Q S \Phi - \Phi S Q)) \frac{2}{\hbar^{2}C} \mbox{ Im}(W_{\rho J}^{+})\ \ \ \ \  \ \ \ \ .
\end{eqnarray}

It is worth noting that this equation is very clean: the coefficients of the various terms do not depend on (nor were they chosen to match)  any  classical/quantum criteria that is outside the scope of the Hamiltonian, and furthermore  the coefficients are directly related to Green's functions of the current and charge density operators.   There are  more terms than expected.   But more importantly, the physics contained in  those terms is clear and fundamental.   The terms proportional to  $\mbox{ Im}(W_{\rho}^{+})$, and  $ \mbox{ Im (W}_{J}^{+}) $ yield the change in the density matrix due to the change in the rest mass of the particle.   Examination of $\mbox{ Im (W}_{J}^{+}) $ shows that it is similar to the 
calculation of the renormalized mass of a free electron that is interacting with a radiation field.

The  term $ L_{a}=[Q,[Q,S]] \alpha_{L}$ , has received a lot of attention, since supposedly it is required   to preserve positivity of the density matrix \cite{diosi}, \cite{munro+gard},  \cite{sgao}, \cite{bvacc}.   And  despite significant efforts $( \cite{diosi}, \cite{sgao}, \cite{bvacc} )$,  the coefficient   of that term has only been obtained at  high temperature $( \alpha_{L} = \alpha_{L1}/( k_{B}T )$,  thereby leaving the question of positivity still open at   low temperatures.  Furthermore the constant     $\alpha_{L1}$ is not the same  in the different works.    It was shown by Lindblad that positivity of $\rholg(t)$ is assured for  Master equations of the Lax-Louisell form: \ $ \dot{\rholg}= -i\hbar^{-1}[H,\rholg] \, + \, \sum_{j}( [A_{j}^{\dagger},\rholg A_{j}] + H.C)$  (where $A_{j}$ is composed of system operators). 
  Under the conditions implicitly  considered by the aforementioned investigators (high temperature   and  the dissipative medium is not quantized),  equation (~\ref{ms1}) is of the Lax-Louisell form.  Under all conditions equation (~\ref{S4}) is of the Lax-Louisell form, and  (~\ref{ms1}) would have been of that form if   terms proportional to the real part of the cross correlation were kept. 
  However when  the correlation functions are   computed  for the dissipative medium in equilibrium, then Re$W_{\rho J}^{+} =0$.  If the dissipative medium is driven very far from equilibrium then the aforementioned terms can be reinserted, and the Re$W_{\rho J}^{+} $ computed using equation (~\ref{drdt2}).
The really astonishing result concerning $L_{a}$ that is reported here 
 is that  the coefficient ($ \alpha_{L}$)  is related
to an important physical quantity: the longitudinal dielectric function (or the density susceptibility function when the system operators are position and momentum).  
Hence with (~\ref{dencor1}) it   is now possible to give  an  exact expression for ($ \alpha_{L}$), that is valid at all temperatures: 
\begin{equation}
\label{dencor2}
\mbox{Re}(\, W_{\rho}^{+}(t) \,) \ = \ - \frac{\hbar}{(2 \pi)^{2}}V_{ol} \ \lim_{\qv \rightarrow 0}    \ \int_{-\infty}^{\infty} \frac{ |\qv|^{2} }{1 - e^{-\beta \hbar \omega}} \  \mbox{Im}\left (\frac{1}{\epsilon_{\parallel}(\qv , \omega)} \right ) \,  \frac{\sin(\omega t)}{\omega} \    d \omega \ \ \ \ \ \  \ \ .
\end{equation}
The physical mechanism associated with $L_{a}$  is  the screening of  coherent charge fluctuation (on the capacitor) by the medium.  One can say that the friction ($R(\omega)$) damps the ensemble-averaged charge/momentum, while the coherent momentum fluctuation is damped by the density susceptibility.

In a  semiclassical computation, the diagonal elements $\lang u|S|u \rang$  can be    treated as   a  quasiprobability distribution, here $|u>$ is the eigenvector of the charge $Q|u> = u|u>$.   The term $L_{a}$ does not affect evolution of the diagonal elements.  From (~\ref{ms1}), the evolution of $<u|\dot{S}|u>$  is given by  a  Fokker-Planck equation,  with a probability diffusion coefficient $ D_{uu} =   \mbox{ Re }( W_{J}^{+} )$.    In the purely ohmic limit   ($\tau_{m} \rightarrow \delta$),   \ \   $ D_{uu} \simeq  K_{B}T/R_{s} + \hbar/(\pi R_{s}t)$, and  hence for all times $t < \hbar \beta / \pi $, the diffusion and particle's evolution are purely quantum mechanical.   
Evolution of the ensemble average \ $<Q> \equiv Tr\{QS(t)\}$, is given by
\begin{equation}
\label{dQdtcl}
 \frac{d <Q>}{dt} -\frac{2}{\hbar C} \mbox{Im}(W_{J1}^{+}) <Q> = I_{s}
 \end{equation}
which reduces to the classical equation: $\dot{Q} + (R_{s}C)^{-1} Q = I_{s}$.  
  Equation (~\ref{dQdtcl}) represents a true milestone: the \underline{derivation} ---for the first time--- of the classical equation that describe a dissipative system from a purely quantum theory.  Note that  $W_{J1}^{+}$ is directly related to $\psi_{d}$, the wavefunction of the dissipative medium.  In theories that use a collection of harmonic oscillators, the classical equation is not derived, but rather it is used to fit the free parameters (spectrum, etc) of the theory.  \hspace{\fill} $---<$np5$>$

It is now possible to write ---for the first time---  a reduced density matrix that describes  the evolution of a dissipative medium that is coupled to a quantum object. Continuing from (~\ref{drdt1}), I obtain:
\begin{eqnarray}
\label{drdt2}
 \frac{ \partial \rholg_{d} }{\partial t}  & =  &    \frac{1}{i \hbar} \left [ \, H_{d} + W_{d} +  V_{d} \, , \, \rholg_{d} \, \right ]  \  - \    \lcommb  I_{d} \, , [ I_{d} \, , \, \rholg_{d} ] \rcommb \frac{1}{\hbar^{2}} \mbox{Re} \, W_{\Phi \Phi}^{+}(t)     
  \nonumber \\
 & &   +   \ \left ( \, [ Q_{d} \, , \{ I_{d} \, , \, \rholg_{d} \} ] -  [ I_{d} \, , \{ Q_{d} \, , \rholg_{d} \} ] \, \right ) \frac{i}{\hbar^{2}} \mbox{Im} \, W_{Q \Phi}^{+}(t)   -    \lcommb  Q_{d} \, , [ Q_{d} \, , \, \rholg_{d} ] \rcommb \frac{1}{\hbar^{2}} \, W_{QQ}^{+}(t)  \nonumber \\
& & \ + \  \lcommb  I_{d} \, , \{ I_{d} \, , \rholg_{d} \} \rcommb \frac{i}{\hbar^{2}} \mbox{Im} \, W_{\Phi \Phi}^{+}(t)  
  \ \ \ \ \ \ \ \ \ \ \  . 
\end{eqnarray}

 The charge operator associated with the charge density is defined as: \  $ Q_{d} \, = \, \lim_{\qv \rightarrow 0} \rholg_{cd}(\qv) $, and  $I_{d}$ (amps)  is the long wavelength  limit of the current through the resistor.  In this  bistratum quantization approach, 
 one set of operators    describe  the individual modes or individual particles of the   dissipative medium ($C_{\lambda}'s$, see numbered paragraph 2 (or$\lang$ np2 $\rang$)), while the key operators that describe the coupling and enter the Master equations  1) satisfy a commutation relationship in wavevector space, and 2) are composed of the microscopic operators.     Equation (~\ref{drdt2}) still contains the full $H_{d}$  and hence a direct solution is far from trivial.   Nonetheless, the equation is manageable because the results of the  equilibrium (only $H_{d}$) solution are known.   The partial spectral functions are given by:
$$\label{wpp}
W_{\ensuremath{\Lambda_{i} \Lambda_{j}} }^{+}(T,t) =  \  \int_{0}^{t} \ dt' \ Tr_{s} (S(t') \Lambda_{i}(t-t') \Lambda_{j})\ - \  
Tr_{s} (S(t') \Lambda_{j}) Tr_{s} (S(t') \Lambda_{i}(t-t') ) 
$$ 
with $\Lambda_{1}=\Phi$, and $\Lambda_{2}=Q/C \ $.

Equations (~\ref{ms1}) and (~\ref{drdt2}) radiate the sublime grace of  the new method: bistratum quantization, in concert with a  coupled systems approach  that recognizes
 both objects as quantum entities.  The foundation of  theories of  dissipative quantum systems has so far been inscrutable.  In contrast, here the quantization of the medium,  quantization of the    operators used to construct the interaction Hamiltonian,  and the  fundamental connection between those operators and  the microscopic operators are all clear. Parenthetically, it is possible to  draw a distinction between an open system (typically found in quantum optics), and  an inherently  dissipative system. In an open system such as an atom in a cavity, a process such as  spontaneous emission is sometimes viewed as dissipative,  but if the number of modes is reduced then the process becomes reversible. In contrast,  resistance to  even dc current flow,  and irreversibility  is present  in the typical  closed system. The difference is due to the underlying mechanisms in the dissipative medium: in one case a set of independent modes, and in the other a strongly coupled ($H_{ee}$) electron system.   It is important to note that an intrinsically dissipative medium   is not really quantized in models that employ a collection of oscillators (or similar).    One may think that   under the quantization of the  harmonic oscillators, ...  but quantization in a space is irreducible,  there is  nothing under, .... under.  A  direct consequence being that the evolution of the dissipative medium  could not  be determined, and hence questions concerning  operators of that medium were part of the unspeakable.

In summary, equations for the reduced density matrices of two interacting sub-systems, a free particle and a linearly dissipative medium,  were derived. The approach to the 40 year conundrum  (see $<$np1$>$ ) of  a  Hamiltonian description of the interacting sub-systems,  and quantizing the whole  system is clear.  
  Operators that satisfy a commutation relation  ---in wavevector space--- were utilized  to construct an interaction   Hamiltonian.  
 The total density matrix was written as a tensor product of the reduced density matrices plus a correlation matrix.   Partial spectral densities that appear in the Master equation for the particle  are given in terms of well known quantities: the retarded current-current correlation function, and the retarded density-density correlation function.

The theory  is particularly important because  a) in many cases of interest,   operators belonging to the dissipative medium  carry the required information and now questions that contain those operators can be answered, and b)    it appears to be   the basis  of a truly quantum circuit theory.  For example, in  the system shown in figure 1 (which may be part of a larger circuit),  the mean current through the resistor, and the fluctuation of that current often  correspond to the detected signal and the fluctuation of the signal.    A circuit theory that can describe quantum  transport is particularly important these days, because everybody's vision of the future {\em now} includes nanoscale and macroscopic quantum circuits. 

One of the most  surprising  features of Brownian motion that was unveiled in this investigation  is the role played by the longitudinal dielectric function.   It is easy to understand why all previous investigations  have missed this important feature. Classical and semiclassical studies are based primarily on  the Langevin, or quantum Langevin equation (for the momentum), and as can be ascertained from $\lang np5 \rang$,  the density correlation does not contribute to the Langevin noise force.   Quantum  studies were based on reproducing  the friction constant $R(\omega)$, either  directly
through the Caldeira-Leggett model,  or indirectly through  a fluctuating force and the fluctuation-dissipation theorem.  However as we have seen the dynamics of the medium is underdetermined  by $R(\omega)$, or the associated fluctuation-dissipation theorem. How reliable are seemingly exact  quantum models  of an  object, where exactness is determined by some classical criteria? It appears that  a quantum object can only be known   via the correct  Hamiltonian.

\newpage


\end{document}